\documentclass[a4paper]{jpconf} % letterpaper could be used instead of a4paper
\usepackage{graphicx}
\usepackage[noadjust]{cite}
\usepackage{amsmath,amsfonts,amssymb}
\usepackage{graphicx}
\usepackage{subfig}
\usepackage{float}
\begin{document}

\title{A new approach to the experiment intended to test the weak equivalence principle for the neutron}

\author{M.A. Zakharov, G.V. Kulin, A.I. Frank, D.V. Kustov, S.V. Goryunov}

\address{Frank Laboratory of Neutron Physics, Joint Institute for Nuclear Research, Dubna, Russia}

\ead{kulin@nf.jinr.ru}

\begin{abstract}
A new approach to the free fall experiment with UCN is proposed. The idea is that the experiment is performed with neutrons from different equidistant lines of a discrete energy spectrum with a precisely known distance between the lines. Such a spectrum may be formed by neutron diffraction from a moving phase grating. Time-of-flight values of neutrons from different lines of the spectrum must be measured. Neither the initial neutron energy, nor the geometry parameters of the installation are required to be known in this method.
\end{abstract}

\section{Introduction}
Apparently, neutrons are the most suitable objects to investigate the gravity interaction of elementary particles. Although gravitational experiments with neutrons have a more than half a century history, the existing experimental data are quite scanty. Almost fifteen years after the first observation of the
neutron fall in the Earth’s gravitational field \cite{PhysRev83}, the gravitational acceleration neutron was measured in a classical experiment with an accuracy of about 0.5$\%$ \cite{PhysRev139}. However, the fact of gravitational acceleration of the neutron had already been considered earlier as obvious and used by Maier-Leibnitz and Koester  \cite{Physik14,ZPhys182} for precise measurements of the coherent scattering length of neutrons by nuclei. 

Later on, there appeared data on coherent scattering lengths for neutron–nucleus interaction that were obtained by measuring cross sections for neutron scattering on atoms. This made it possible to compare scattering length data obtained by two methods and thereby to verify the fundamental
principle of the equivalence of the inertial and gravitational masses of the neutron \cite{PhysRevD14}. He obtained the value $\gamma$ = 1.0016 $\pm$ 0.00025 for an equivalence factor that he defined as $\gamma=(m_{i}/m_{g})/(g_{n}/g_{0})$, where $m_{i}$ and $m_{g}$ are the inertial and gravitational neutron masses, respectively, and $g_{n}$ and $g_{0}$ are the gravitational acceleration of the neutron and the local gravitational acceleration of macroscopic bodies, respectively. More recently, a similar analysis was performed by Schmiedmayer \cite{NucInstrMethA284}, who
obtained the equivalence factor with an accuracy twice as good as that obtained in \cite{PhysRevD14}. 

The first quantum neutron gravitational experiment was performed in 1975 by Colella, Overhauser, and Werner \cite{PhysRevLett34} with a neutron interferometer. They observed the gravitation induced shift of the phase
of the neutron wave function. Results of first experiments \cite{PhysRevLett34,PhysRevA21} basically corresponded to theoretical predictions. However, further investigations revealed certain discrepancies. In the latest work \cite{PhysRevA56}, the difference between the experimental and theoretical phase shifts was equal to 1$\%$ with an error smaller by an order of magnitude. Results of a more recent experiment \cite{NuclInstrMethA440}, the accuracy of which was equal to 0.9$\%$ do not remove this problem. Very recently, an experiment with a neutron spin-echo spectrometer has been performed \cite{PhysRevA89}. The authors reported that their experimental result for the gravitation induced phase shift agrees within approximately 0.1$\%$ with the theoretically expected result, while the overall measurement accuracy was 0.25$\%$.

Nesvizhevsky et al. \cite{Nature415,PhysRevD67} reported the observation of the quantization of the vertical-motion energy of ultracold neutrons (UCNs) stored on a horizontal mirror. It is possible to hope that detailed investigation of this effect will be very useful for studying the gravitational interaction of the neutron as a quantum particle. Very promising results have been recently obtained by Jenke et al. \cite{NaturePhysics7}, who observed transitions between quantum states of UCN being stored on a plane mirror in the Earth’s gravitational field.

In the experiment \cite{JETPLetters86} the change in the energy of a neutron falling to a known height in the Earth’s gravitational field was compensated by a quantum of energy $\hbar \Omega$ that was transferred to the neutron via a non-stationary interaction with a moving diffraction grating. In this experiment the value of the equivalence factor defined as $\gamma=m_{i} g_{n}/m_{n} g_{loc}$ was found to be $1-\gamma = (1.8 \pm 2.1)×10^{-3}$. Here $m_{i}$ is neutron inertial mass, $m_{n}$ and $g_{loc}$ are neutron table mass and local value of the free fall acceleration, respectively.

In the present paper we discuss the possibility of a new type of experiment for testing the weak equivalence principle for the neutron.

\section{New approach to the free fall experiment with the neutron}
Let us assume that the weak principle of equivalence is tested in a classical type of the experiment called free-fall or Galileo type of experiments. In such an experiment it is necessary to define the parameters of the equation of the neutron motion at its free falling in the Earth’s gravitational field. It
is obvious that in order to measure the free-fall acceleration for the neutron it is necessary to know the time of flight, initial velocity and height of falling with an appropriate accuracy.

The original approach to the free fall experiment was proposed by Pokotilovsky \cite{PhysAtomNucl57}. He proposed to use the geometry of a “neutron fountain”, in which ultracold neutrons (UCN) following a parabolic trajectory move up and, after passing the turning point, fall down on the annular detector.

In the experiment the time of flight along such a trajectory is measured
\begin{equation}
t=\frac{v_{0}}{g_{n}}\left( \sqrt{1+\frac{2 g_{n}h}{v_{0}^{2}}}+1\right).
\label{eq:classicTOF}
\end{equation}
Here $g_{n}$ is the neutron free-fall acceleration, $v_{0}$ is the vertical component of the initial velocity and h is a difference in height between the source and detector. The most challenging in this approach is measurement of the neutron vertical velocity with high precision. It was proposed to circumvent this problem by a repetitive measurement of time with a set of the heights $h_{k}$. It multiplies the number of equations, from the solutions to which the value of $g_{n}$ may be found. For implementation of this experiment it is necessary to vary the position of the detector in height, thus ensuring stability of its horizontal orientation with high accuracy.

\begin{figure}[H]
	\begin{minipage}[t]{0.45\textwidth}
		\includegraphics[width=\textwidth]{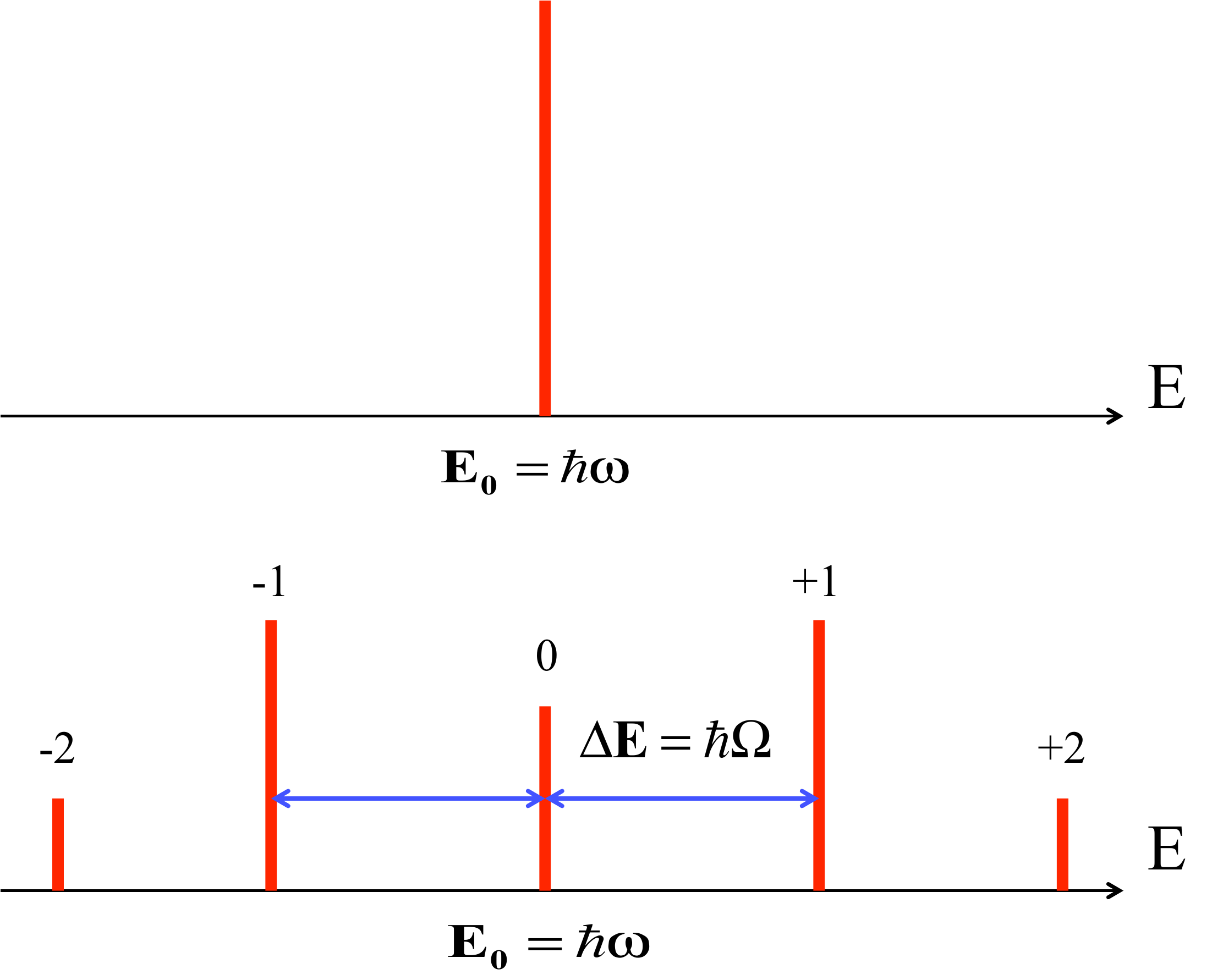}
	\end{minipage}\hspace{0.034\linewidth}
	\begin{minipage}[b]{0.5\textwidth}
		\caption{Splitting of the energy spectrum by a non-stationary quantum device.} \label{fig:Esplitting}
	\end{minipage}
\end{figure}

We propose an alternative approach to the Neutron Fountain experiment, which permits avoiding the necessity of the mechanical displacement of any elements of the installation. The idea is to use any quantum non-stationary device, which transforms the initially quasi-monochromatic spectrum to the discrete spectrum with an exactly known value of splitting $\Delta E = \hbar \Omega$, as it is shown in Figure \ref{fig:Esplitting}. The time-of-flight values of neutrons with a set of energies corresponding to different spectral lines must be measured simultaneously.

The free-fall acceleration $g_{n}$ can be found from the system of equations

\begin{equation}
t_{j}=\frac{1}{g_{n}}\sqrt{\frac{2(E_{0}+j\hbar\Omega)}{m_{i}}}\left[ \sqrt{1+\frac{m_{i} g_{n} h}{E_{0}+j\hbar\Omega}}+1\right] ,
\label{eq:classicTOFsystem}
\end{equation}
where $E_{0}$ is the initial energy of the vertical motion and $j$ is the order of a satellite line (see Figure \ref{fig:Esplitting}). The number of equations may be increased further by repetition of measurement with a set of frequencies $\Omega_{k}$.

\section{Possibility of experimental implementation}
As it is proposed in \cite{PhysAtomNucl57}, the experiment must be performed with UCN. The quasi-monochromatic beam will be formed by neutron transmission through the quantum resonant structure - Neutron Interference Filter (NIF) \cite{JETP,PhysRevLett44,ProcSPIE3767}, which is analogous to the optical Fabry-Perot interferometer. In order to transform the quasi monochromatic spectrum into the spectrum with equidistant lines, a moving diffraction grating \cite{PhysLettA311,JETPLetters81,PhysRevA2016} will be used. The time-of-flight values of neutrons from different diffraction orders will be measures simultaneously by the Fourier spectroscopy method \cite{NuclInstrMeth2016}.

\begin{figure}[H]
	\centering
	\includegraphics[scale=0.31]{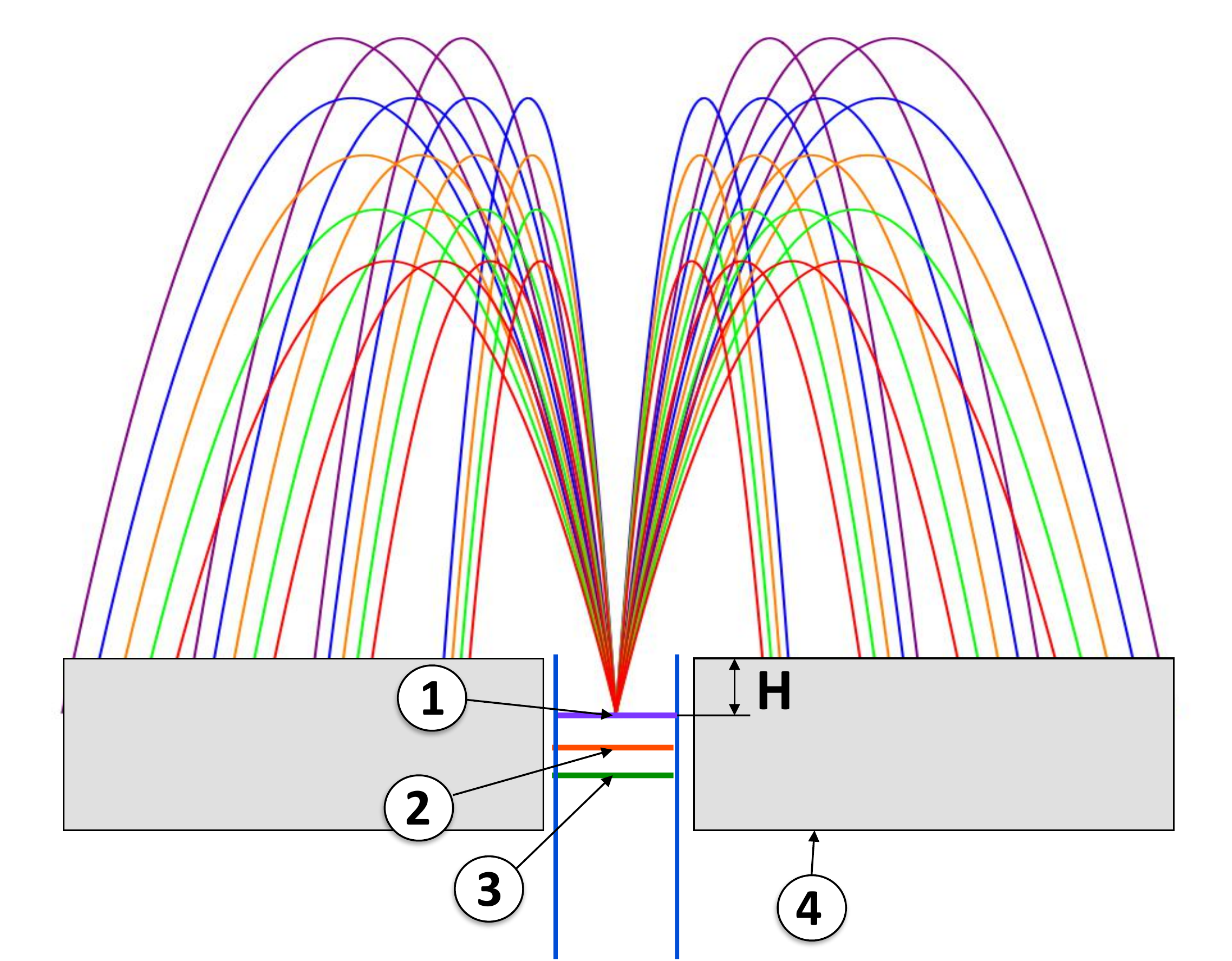}
	\caption{Scheme of the experiment: 1 - neutron interference filter (NIF), 2 - moving diffraction grating, 3 - Fourier chopper, 4 - annular detector.}
	\label{fig:ExpScheme}
\end{figure}

The frequency of the wave quantum modulation is $\Omega = 2\pi f$ , $f =V d$, where $V$ is the linear speed of the grating and $d$ is its space period. The scheme of the proposed experiment is shown in Figure \ref{fig:ExpScheme}. UCNs from the feeding neutron guide pass sequentially through the NIF, moving grating and Fourier chopper, and after that, moving along a parabolic trajectory, fall down on the annular detector.

\begin{figure}[H]
	\begin{minipage}[t]{.48\textwidth}
		\includegraphics[width=\linewidth]{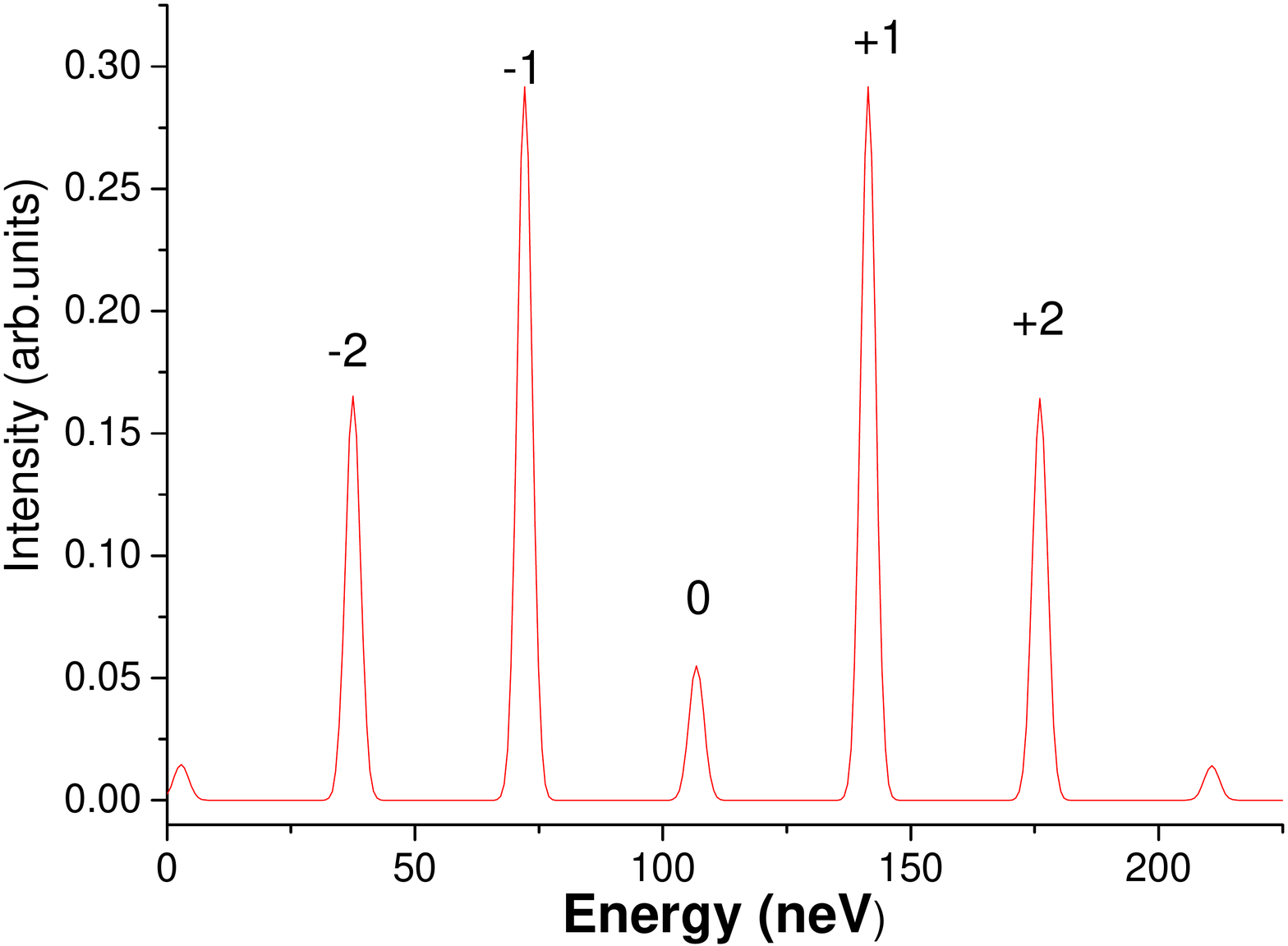}
		\captionof{figure}{Calculated spectrum of UCN with initial energy of 107 neV after diffraction from the moving grating. Modulation frequency $f$ =	8.4 MHz.}
		\label{fig:CalcSpectrum}
	\end{minipage}\hspace{0.03\linewidth}
	\begin{minipage}[t]{.48\textwidth}
		\includegraphics[width=\linewidth]{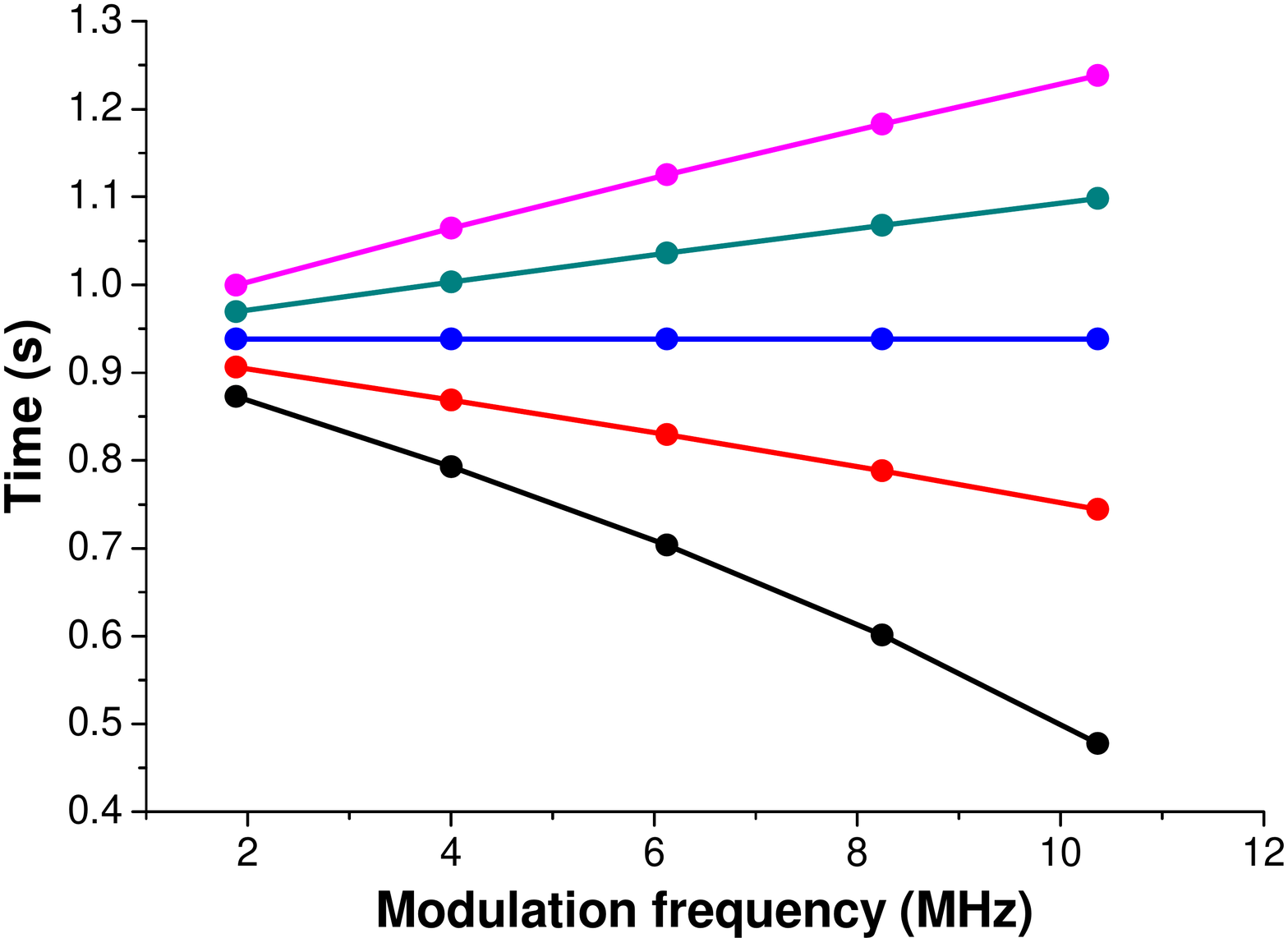}
		\captionof{figure}{Time of flight of neutrons of the five diffraction orders in dependence on modulation frequency.}
		\label{fig:TOFfiveorders}
	\end{minipage}
\end{figure}

To estimate the relation between the accuracy of the measured free-fall acceleration gn and time of flight experimental errors, a model calculation was done. It was supposed that the spectrum of UCN formed both by NIF with a resonant energy of 110 neV and a moving grating operating with the frequency $f$. As an illustration, the calculated spectrum formed by a specially designed grating with increased intensities of the high orders \cite{BushuevarXivJETP2015} is shown in Figure \ref{fig:CalcSpectrum}. The time-of-flight values of neutrons with energies corresponding to the five spectral lines were calculated in dependence on the modulation frequency for the specified values of $g_{n}$, $E$ and $h$ (see Figure \ref{fig:TOFfiveorders}). For the five values of modulation frequencies pseudo-experimental data for the time of flight were found. It was supposed that the latter were defined by normal distribution with dispersion corresponding to the specified errors in time measurement. Twenty five values of the time of flight found using this procedure (see points in Figure \ref{fig:TOFfiveorders}) were fitted by the functions
\begin{equation}
t_{j,k}(g_{n},E_{0},h)=\frac{1}{g_{n}}\sqrt{\frac{2(E_{0}+j\hbar\Omega_{k})}{m_{i}}}\left[ \sqrt{1+\frac{m_{i} g_{n} h}{E_{0}+j\hbar\Omega_{k}}}+1\right] .
\label{eq:classicTOFfitfuncs}
\end{equation}

As a result, all three unknown parameters, namely $g_{n}$, $E_{0}$ and $h$, were found with the corresponding errors. These calculations show that the error of $g_{n}$ is proportional to the errors of the time of flight and relative accuracy of $\delta g_{n}/g_{n}\approx 10^{-5}$ − may be achieved, if the errors of the time measurement are of an order of 10mks.

\section{Conclusions}
A new approach to the free fall experiment with UCN is proposed. The idea is that the experiment is performed with neutrons from different equidistant lines of a discrete energy spectrum with a precisely known distance between the lines. Such a spectrum may be formed by neutron diffraction from a moving phase grating. Time-of-flight values of neutrons from different lines of the spectrum must be measured. Neither the initial neutron energy, nor the geometry parameters of the installation are required to be known in this method.

\ack
This work was supported by the Russian Foundation for Basic Research (RFBR grant 15-02-02509).
 
\section*{References}
\bibliography{ref}

\end{document}